\documentclass[numreferences]{kluwer}    
\newdisplay{guess}{Conjecture}
\usepackage{psfig}
\usepackage{klucite}
\usepackage{journals}
\begin{document}

\begin{article}

\begin{opening}         
\title{Exploring the Jet/Accretion Flow Relationship in Low Disk
            Luminosity Sources\thanks{NSF Astronomy \& Astrophysics Postdoctoral Fellow}} 
\author{Sera \surname{Markoff}}
\runningauthor{Sera Markoff}
\runningtitle{Jets/Accretion in Low Luminosity Sources}
\institute{MIT, Center for Space Research, Cambridge, MA 02139, USA}

\begin{abstract}
Astrophysical jets seem to gain strength disproportionate to the power
of their associated accretion flow, making low-luminosity sources
ideal targets for studies of the role of outflows.  Radio/X-ray
correlations have supported the case for a strong relationship between
the jets and the hard X-ray emitting regions, and here we explore the
strongest scenario where the base of the jets subsumes the role of the
corona.  The properties of coronae, as inferred from spectral models,
are very similar to what is empirically required at jet bases assuming
conservation laws hold.  We present a few preliminary fits to
simultaneous radio and X-ray data sets from our GX~339$-$4 and Cyg X-1
campaigns.  The fits are performed in detector space, and include a
jet plus thermal disk continuum model, with added Gaussian line and
non-relativistic reflection features similar to the approach of other
X-ray models.  We find that we can fit the entire radio through X-ray
spectrum quite well, with any deviation occurring in the
line/reflection region.  The results suggest that a jet/corona
unification can provide a reasonable description of the data.  Future
work will benefit from a more complex approach to the disk feedback
features.

\end{abstract}

\end{opening}           

\section{Introduction}  

Relativistic outflows have been considered part of the big picture for
active galactic nuclei (AGN) almost since their discovery, simply
because their scales are so large compared to the host galaxy in radio
frequencies.  The confirmation that jets also exist in smaller
accreting compact objects, such as X-ray binaries (XRBs), was not
fully established until the late 1970s (SS433; \cite{Spencer1979}).
However, XRB jets continued to be treated as distinct components until
fairly recently, when correlations between the radio and X-ray
luminosities were discovered in the hard state of GX~339$-$4
\cite{Hannikainenetal1998,Corbeletal2000,Corbeletal2003}.  This
correlation is now thought to be fundamental to accreting black holes
\cite{GalloFenderPooley2003,MerloniHeinzDiMatteo2003,FalckeKoerdingMarkoff2004}.
The realization that the correlation scales with mass has provided a
new method of comparison between AGN and XRBs, and for the
exploration of the spectral roles of the individual components.

Despite this progress, the details of the inflow/outflow interface are
still a matter of significant debate.  The nature of the connection is
important for models of jet formation, and in general our
understanding of the effects of strong gravitational fields on
magneto-hydrodynamical plasmas.  An important way to approach this
problem is to create physical models which consider the system
holistically, by both predicting the broadband spectra at the same
time as addressing fine features specific to the X-ray band.  Since
the radio emission in these sources is accepted to come from jets we
have a solid anchor for models which probe how jets relate to the
X-ray emission.

Low disk luminosity systems are characterized by their sub-Eddington
accretion rates and steady compact jets. They are advantageous sources
for this study because one can avoid the complications of high
accretion rates as seen in, e.g., AGN, especially near the base of the
jets where the physics is most in question.  Furthermore, because the
evidence is increasing that jets dominate the power at low
luminosities
(e.g., \cite{FenderGalloJonker2003,MalzacMerloniFabian2004})
we can
better hope to isolate the roles of jets in these sources.

\section{Model Background}

\subsection{Jet model development}

We first explored scaling jet models for Sgr A*, whose quiescent and
flared emission can be explained via synchrotron radiation in the
radio frequencies, and synchrotron and/or synchrotron self-Compton
(SSC) in the X-ray band \cite{FalckeMarkoff2000,Markoffetal2001}.
This model scales with the central mass and accretion power, and thus
in principle can be applied to XRBs as well.  XTE~J1118+480 was the
ideal test source, with its unprecedented high-quality broadband,
quasi-simultaneous data (\cite{Hynesetal2000,McClintocketal2001b}, and
refs. therein).  We had originally set out to model the radio spectrum
via synchrotron emission, but it was immediately apparent that the
scaling resulted in the optically thick-to-thin break occurring in the
infrared/optical bands.  More importantly, we found that if a power
law distribution of energetic particles exists further out in the jet,
it is in fact quite difficult to suppress the optically thin
synchrotron radiation from extending out into the X-ray band
\cite{MarkoffFalckeFender2001}.  An accelerated power law distribution
of particles is motivated by observations of optically thin emission
during radio outbursts in XRBs (e.g., \cite{FenderKuulkers2001}) and
AGN (e.g., \cite{MarscherGear1985}).  The jet synchrotron-dominated
model presented in \cite{MarkoffFalckeFender2001} gives a good
description of the broad spectral features of XTE~J1118+480, with the
$\sim 100$ keV cutoff following from acceleration saturated by
synchrotron cooling.  Thermal emission from the inner edge of an
accretion disk was included in the calculation both as a weak direct
component and as seed photons for inverse Compton scattering, but was
not the dominant emission component.

We next considered this model for 13 (quasi)-simultaneous radio/X-ray
(and sometimes IR) data sets for GX~339$-$4 \cite{Corbeletal2003}.  We
found again that jet synchrotron could explain the broadband continuum
spectra of all the observations, mainly by just varying the power
input into the jet \cite{Markoffetal2003}.  We showed in this paper
that jet synchrotron emission analytically predicts the slope of the
radio/X-ray correlation as a consequence of its scaling with power.   

In summary, these simple synchrotron-dominated jet models
significantly developed our understanding of the radio/X-ray
correlations and scaling.  However, they did not attempt to address
the fine features in the X-ray spectrum, which are harbingers of
interactions with cooler accretion disk material.

\subsection{Jet/corona relationship}

Accounting for the hard state X-ray power law spectrum, in combination
with the line emission/reflection features attributed to interaction
with the accretion disk, led to the development of the corona/disk
model (e.g.,\cite{HaardtMaraschi1991,GeorgeFabian1991}).  The corona
is inferred from X-ray signatures to have several properties, such as
quasi-thermal electrons in a reasonably compact geometry.  The
fraction of reflected hard X-ray emission is often low in the hard
state, and this fact is sometimes problematic for corona models
\cite{dove:97b}.  Various mechanisms have been proposed to decrease
the fraction of reflected X-rays, including patchy coronae
\cite{stern:95b}, high disk ionization
\cite{ross:99a,Nayakshin2000,ballantyne:01a} and beaming of the
coronae away from the disk with mildly relativistic velocities
\cite{Beloborodov1999,MalzacBeloborodovPoutanen2001}.  This latter
approach seems extremely close in principle to the characteristics of
the base of a jet.

Unlike corona jets are observed explicitly: the synchrotron emission
tells us that there is a population of accelerated leptons quite far
out in the jet.  If one believes that there is rough conservation of
properties along the jet, then tracing back to the base requires
hotter leptons in even denser populations threaded by even stronger
magnetic fields.  At the same time, magnetohydrodynamical simulations
of the inner disk region do not show anything like a stable
sphere/disk geometry, but rather a region threaded with fields which
naturally leads to outflowing plasma (corona?) (e.g.,
\cite{StonePringle2001}).  This beaming will reduce
the relevance of the already weak thermal disk photons with respect to
locally created photons and lead to an altogether different picture
than the static case.  We think it is worth exploring the most
extreme, and thus easiest to test, case scenario: whether the base of
the jets can ``subsume'' the role of the corona.  This is a critical
step in the process of disentangling their actual relationship, since
if the approach fails, where and how it fails will provide clues as to
the nature of the corona as a distinct component.  If it does not
fail, it may help point the way towards a better understanding
of jet formation.

\subsection{Model Background}

\begin{figure}
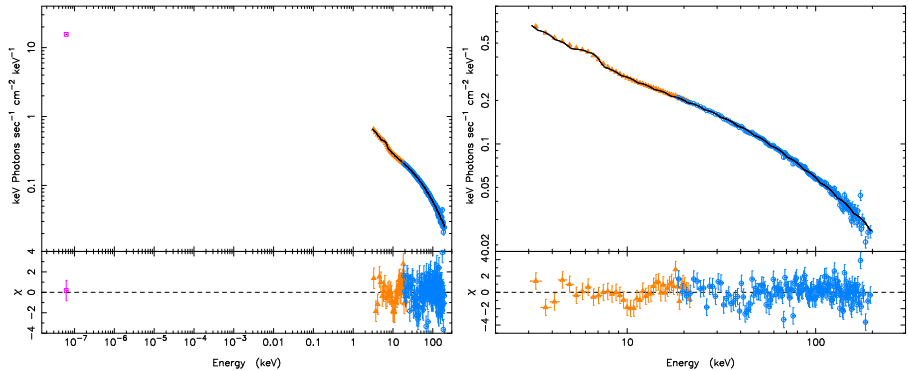

\hbox{
\psfig{figure=fit1.14_radio.ps,width=0.5\textwidth}
\psfig{figure=fit1.14_xray.ps,width=0.5\textwidth}
}
\caption{The best fit so far of our three Cyg X-1 simultaneous
radio/X-ray observations which span its typical spectral variations in
the hard state, $\chi^2=208.4/183=1.14$.  The figures show the
unfolded fit plus residuals with an absorbed jet model (including a
thermal accretion disk component) plus Gaussian line, convolved with
non-relativistic reflection.  a) The entire broadband fit, including
the radio data point, b) close-up on the X-ray bands.  X-ray data are
from RXTE PCA \& HEXTE, radio from the Ryle Telescope.}\label{cygbest}
\end{figure}

A more detailed explanation can be found in other papers
(\cite{MarkoffFalckeFender2001,Markoffetal2003}; Markoff, Nowak \&
Wilms in prep.) but the basic model is a freely expanding jet which
has a velocity gradient due to acceleration along its axis.  Plasma
enters the jet at the base, and the radiation
is dominated by quasi-thermal leptons which cool radiatively and
adiabatically as the jet expands.  Further out in the jet the
particles encounter an acceleration region where some of them are
accelerated into a power law tail.  The particles radiate along the
entire jet via synchrotron, synchrotron self-Compton (SSC) and
external Compton (EC) radiation.  The EC only contributes very close
to the base of the jet, where it can be comparable to the SSC, both of
which contribute to the hard X-rays.  Synchrotron emission from $\sim
100$ $r_{\rm g}$ dominates in the radio/IR bands, turning over to
contribute in the X-ray band.

Our currently favored model is a direct result of our work in
\cite{MarkoffNowak2004}, where we calculated the reflection from
typical jet models based on GX~339$-$4 spectra.  We found this a very
useful probe of the geometry of jet emitting regions, which
helped define our latest models.  If the acceleration region really
exists further out in the jet, as we have found for several fits, then
the weakly beamed synchrotron radiation cannot result in more than a
few percent reflection (assuming perpendicular geometry and a flat
disk).  SSC near the base of the jets, however, can easily give
$\lesssim20$\% in the simplest case.  

In order to address the fine features of the X-ray spectrum in a
statistical manner, we needed to import this model into X-ray data
analysis software.  Currently it is running in {\tt XSPEC} and {\tt
ISIS}, and the figures here were made with the latter program.  In
comparison to models which focus exclusively on the X-ray frequencies,
however, we are also importing and fitting the simultaneous radio data
with our model.  Within the program, we added a single Gaussian line
to our continuum model, and allowed it to vary between 6--7 keV.  We
then convolved the entire spectrum with a non-relativistic reflection
model derived from the Greens functions of \cite{Magdziarz1995}.
Because of the complexities of directly calculating the interaction of
jet photons with the accretion disk (see \cite{MarkoffNowak2004}), we
could not self-consistently include these features but include the
strength of the reflection hump as a free parameter.  This is similar
to the general approach chosen by pure Comptonization models
(e.g., \cite{coppi:92a,Poutanen1998}). The full calculation of these
features requires a Monte Carlo approach, which we will consider
elsewhere.

\section{Results}

\begin{figure}
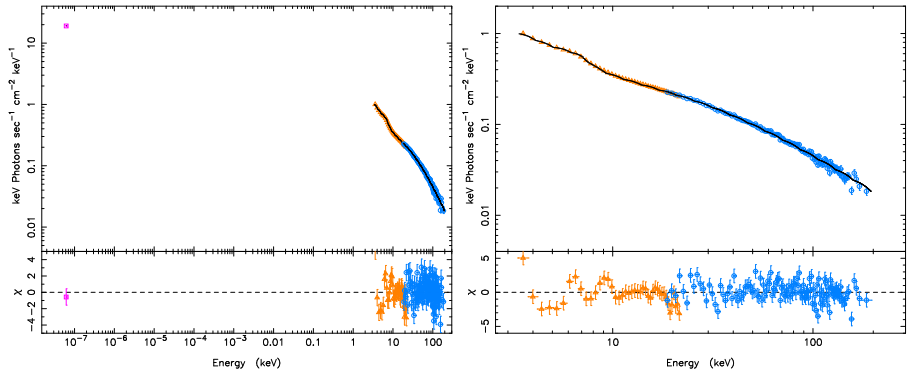

\hbox{
\psfig{figure=fit1.72_radio.ps,width=0.5\textwidth}
\psfig{figure=fit1.72_xray.ps,width=0.5\textwidth}
}
\caption{The worst fit so far of our three Cyg X-1 simultaneous
radio/X-ray observations which span its typical spectral variations in
the hard state, $\chi^2=278/162=1.72$.  Otherwise same format as Fig.~\ref{cygbest}.}\label{cygworst}
\end{figure}

\begin{figure}
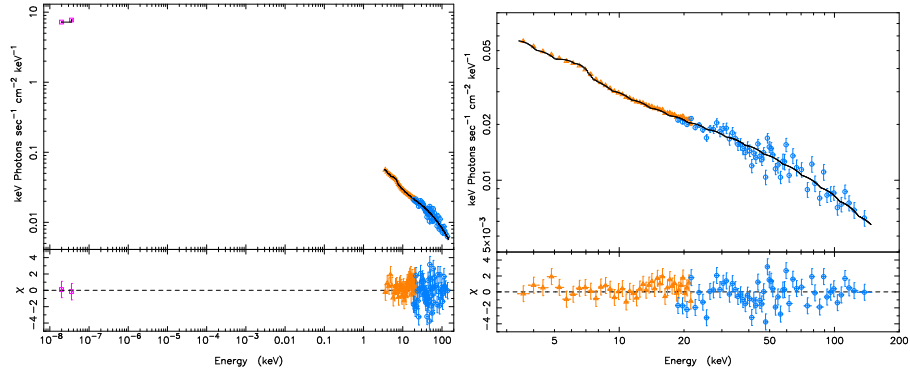

\hbox{
\psfig{figure=fit1.44_radio.ps,width=0.5\textwidth}
\psfig{figure=fit1.44_xray.ps,width=0.5\textwidth}
}
\caption{Typical fit from three simultaneous radio/X-ray observations
of GX~339$-$4 which span its luminosity variations in the hard
state, $\chi^2=138.3/96=1.44$.  Otherwise same format as
Figs.~\ref{cygbest} \& \ref{cygworst}.}\label{gx339}
\end{figure}

Here we present the best and worst fits so far from three data sets of
Cyg X-1 which span its typical range in hard state spectral
characteristics (Figs.\ref{cygbest} \& \ref{cygworst}, respectively).
GX~339$-$4 varies mainly in luminosity rather than spectral shape, and
so we present just one typical fit in Fig.~\ref{gx339}.  The data come
from the Rossi X-ray Timing Explorer (RXTE), using both the
Proportional Counter Array (PCA) and the High Energy X-ray Timing
Experiment (HEXTE; \cite{rothschild:98a}), extracted with the newest
release of the RXTE software, HEASOFT 5.3.1.  The radio data come from
the ATCA (GX~339$-$4) and Ryle (Cyg X-1) instruments.

\section{Discussion}

The fits presented here are comparable to fits of the X-ray
data alone made with broken power law models (Nowak, this volume), and
thermal Comptonization models (e.g. EQPAIR, \cite{Coppi1999}).  Unfortunately,
the multicomponent jet model currently runs significantly slower than
single-component corona models, which means we are not able to explore
the same amount of parameter space in a reasonable amount of time
(yet).  However, it is important to realize that since we are including
the radio self-consistently, we have a more holistic approach to a
system we know is coupled.  The main thing to note is that the
cutoff region is not where the fit has its limitations, but rather in
the line/reflection regime.  This is most likely due to the lack of
time for finding slightly better parameters, as well as the use of
simple, non-relativistic, non-ionized models for the
disk component.  Secondly, the reflection fractions required for
these fits are lower than those inferred for thermal Comptonization
models ($\sim$5--10\% compared to 15--20\%).  This is mainly
due to differences in the shape of the continuum for the two types of models,
because in the jet model the SSC component already lends its curvature
to the hard X-rays, reducing the need for strong
reflection.  At the same time, these lower reflection fractions are
entirely consistent with our estimations of reflection from jet
models, and thus demonstrates that the phenomenology of reflection is
very dependent on assumptions of the impinging continuum.  

So far it seems a model where the jet takes over the role of the
corona, or includes it, is statistically feasible.  The next step is
to consider tighter constraints on the shape of the spectrum at the
highest energies near the cutoff, which may break some of the
degeneracy.  The upcoming mission ASTRO-E2 will hopefully provide
useful data for these tests.  Similarly, timing studies will be
valuable for further constraining the location of the emitting
regions.

\acknowledgements

Some parts of this talk evolved into Markoff, Nowak and Wilms (in
prep., 2005), and so thanks go to my collaborators Michael Nowak and
J\"orn Wilms.  I also would like to thank Heino Falcke, St\'ephane
Corbel and Rob Fender for their collaboration in the earlier
development of this work, as well as Guy Pooley for his persistent
monitoring of Cyg X-1.  Support was provided by an NSF
Astronomy \& Astrophysics postdoctoral fellowship under NSF Award
AST-0201597, as well as NSF grant INT-0233441.


\theendnotes

\end{article}
\end{document}